\documentstyle[preprint,aps,epsf,floats,eqsecnum]{revtex}

\tighten

\begin{document}

\preprint{\vbox{\hbox{UTPT--95-24} \hbox{CMU--HEP95--20}
\hbox{DOE--ER/40682--108} \hbox{JHU--TIPAC--95024}
\hbox{hep-ph/9511454}}}

\title{Phenomenology of the $1/m_Q$ Expansion in \\ Inclusive
$B$ and $D$ Meson Decays}

\author{Adam F.~Falk}
\address{Department of Physics and Astronomy,
The Johns Hopkins University\\
3400 North Charles Street, Baltimore, Maryland 21218 U.S.A.\\
{\tt falk@planck.pha.jhu.edu}}
\author{Michael Luke}
\address{Department of Physics, University of Toronto\\
60 St.~George Street, Toronto, Ontario, Canada M5S 1A7\\
{\tt luke@medb.physics.utoronto.ca}}
\author{Martin J. Savage}
\address{Department of Physics, Carnegie Mellon University\\
Pittsburgh, Pennsylvania 15213 U.S.A.\\
{\tt savage@thepub.phys.cmu.edu}}

\date{February 1996}

\maketitle

\begin{abstract} We apply a recent theoretical analysis of hadronic
observables in inclusive semileptonic heavy hadron decays to the
phenomenology of $B$  and $D$ mesons. Correlated bounds on the
nonperturbative parameters  $\bar\Lambda$ and $\lambda_1$ are derived
by considering  data from $B$ decays and, independently, data from $D$
decays. The two sets of bounds are found to be consistent with each
other. The data from $B$ decays are then are used to
extract a lower limit on the CKM matrix element $|V_{cb}|$. We
address the issue of the convergence of the perturbative expansions
used in the analysis, and compare our bounds on $\bar\Lambda$ and
$\lambda_1$ to lattice and QCD sum rule results.  Finally, we argue that
a comparison of the analyses of $D$ and $D_s$ decays provides
evidence for the applicability of parton-hadron duality in
the semileptonic decay of charmed hadrons.

\end{abstract}


\section{Introduction}

The heavy quark limit of QCD is of enormous practical use, because with it one
may describe a wide variety of heavy hadron decay rates and matrix elements in
terms of a small number of parameters.  These parameters reflect
nonperturbative QCD effects and cannot be computed directly.  Instead, they
must either be modeled or, preferably, be extracted from experimental data.
One of the most important applications of the analysis of inclusive decays is
the determination of the CKM matrix element $|V_{cb}|$ from the process
$B\rightarrow X_c \ell\nu$, which is complementary to the extraction from the
exclusive decay $B\rightarrow D^*\ell\nu$.  The computation involves an
expansion in powers of $1/m_b$, and to ${\cal O}(1/m_b^2)$ there appear three
nonperturbative parameters: $\bar\Lambda$ (or equivalently, the quark mass
$m_b$),
$\lambda_1$, and $\lambda_2$.  While $\lambda_2$ may be extracted directly from
the $B$--$B^*$ mass splitting, $\bar\Lambda$ and $\lambda_1$ are not directly
measurable.   Two approaches currently popular in the literature are
to employ various QCD sum rules to estimate these parameters, and to use an
analysis of inclusive semileptonic $D$ decay to fix one linear combination of
them.  However, each of these methods has a severe disadvantage:  the QCD sum
rule
results are not model-independent consequences of QCD, and the expansions in
$\alpha_s(m_c)$ and $1/m_c$ may or may not work well at the low scales relevant
for $D$ decay~\cite{LSW,bds94}.

In a recent analysis~\cite{FLS}, we calculated the leading perturbative and
nonperturbative contributions to moments of the hadronic energy and invariant
mass spectra
in semileptonic heavy hadron decays.  These predictions are particularly
interesting, because experimental information on invariant masses of the
hadrons produced in these decays may
be derived from the reported branching ratios to  exclusive final states.
Furthermore, they rest on the same theoretical basis
as earlier analyses of other semileptonic quantities such as the decay rate and
the lepton energy spectrum~\cite{history}.

In Ref.~\cite{FLS} we performed some preliminary phenomenology based on this
theoretical
analysis, deriving correlated bounds on the nonperturbative parameters
$\bar\Lambda$ and
$\lambda_1$. In this paper we will develop this phenomenology further,
incorporating
additional data and including in the discussion the semileptonic decays of
charmed mesons.  Our
main conclusions are:
\begin{enumerate}
\item The perturbation series appearing in the analysis are under better
control than had previously been thought.
While the two-loop corrections relating $\bar\Lambda$  and $\lambda_1$ to
the
semileptonic decay rate and to the first moment of the invariant  mass
spectrum in
$B\rightarrow X_c
e\bar\nu$ are large, these corrections partially cancel in the relation
between the
semileptonic  decay rate and the first moment of the invariant mass
spectrum.  The
corresponding perturbation series relating the two physical quantities appears
to
be better behaved.  Using the scale setting technique of Brodsky, Lepage and
Mackenzie \cite{BLM}, we find a BLM scale
$\mu_{\rm BLM}=0.26\,m_b$ for the relation  between the first moment and the
semileptonic
width.  This extends our previous result
\cite{FLS} to the case of finite charm quark mass.
\item When combined with the measured semileptonic width of the $B$, the
moments of the invariant mass spectrum and the measured branching fraction to
excited states yield the constraint
\begin{eqnarray}
   |V_{cb}| > \left[0.040-2.9\times10^{-4}
   \left({\lambda_1\over0.1\,{\rm GeV}^2}\right)\right]
   \left({\tau_B\over1.60\,{\rm ps}}\right)^{-{1\over2}}\,.
\end{eqnarray}
While this is consistent with previous determinations of $|V_{cb}|$ from
inclusive $B$
decays~\cite{Nconf}, this result differs from previous extractions in that it
does not depend on any
assumptions about the size of $\bar\Lambda$, nor on QCD sum rule estimates of
the quark masses.
\item The values of $\bar\Lambda$ and $\lambda_1$ extracted from the
semileptonic decay width and first moment for $D$ and $D_s$ decays
are consistent with those obtained from $B$ decays.  The $B$
results are also consistent with recent lattice extractions of
the $\overline{\rm MS}$ mass $\overline m_b(m_b)$\cite{cgms95,dav95}.
The combined results from $B$ and $D$ decays are inconsistent with the
large negative value of $\lambda_1$
extracted from certain QCD sum rules\cite{BB93a}.
\item The theoretical prediction for the {\it difference\/} of the first
moments of the invariant mass spectrum in $D$ and $D_s$ decays is
well behaved and provides a test of
parton-hadron duality.  The comparison with experimental data is quite
successful, providing additional evidence for the applicability of duality to
the decays of charmed hadrons.
\end{enumerate}

Since our conclusions on $D$ decays disagree significantly with those
presented in
Ref.~\cite{bds94},
it is worth commenting on the discrepancy. The authors of Ref.~\cite{bds94}
used the
extraction of $m_b$ from Ref.~\cite{vol95} along with the QCD sum rules
extraction of
$\lambda_1=-0.6\pm 0.1$ GeV$^2$~\cite{BB93a} to conclude that
$1.25<m_c^{\rm pole}<1.40$.   This results in a semileptonic decay width
for the $D$
meson which is at least a factor of two smaller than observed.   However,
it is difficult
to relate this extraction of the pole mass to physical quantities.
The radiative corrections in the relation between
$m_c^{\rm pole}$ and the semileptonic charm width are so large that the
perturbation series appears uncontrolled; whether or not this is the case for
the
relation between the
moments of $\sigma(e^+e^-\to\bar b\, b)$ (from which $m_b^{\rm pole}$, and
hence
$m_c^{\rm pole}$, are extracted) and the
charm quark semileptonic width requires a higher order calculation.  Given
this uncertainty,
we prefer to treat $\lambda_1$ and $\bar\Lambda$ as free parameters, to be
fixed by
relations between the decay widths, moments and $\overline{\rm MS}$ masses,
in which case we
find that  all the data on charm and bottom are consistent with the
smaller value
$\lambda_1\simeq -0.1\,{\rm GeV}^2$.   This is also consistent with the
observations of Ref.~\cite{Neu94a}, where it is argued that the correct QCD sum
rule should give a substantially smaller value of $\lambda_1$ than that found
in Ref.~\cite{BB93a}.

Finally, we note that we will consider values of $\lambda_1$ which violate the
constraint $\lambda_1\le -3\lambda_2\approx-0.36\,{\rm GeV}^2$ which was
proposed in Ref.~\cite{BSUV95}.  The proof of this bound is criticized in
Ref.~\cite{GKLW}, where it is demonstrated that the inclusion of radiative
corrections precludes any rigorous constraint on $\lambda_1$.  Hence we
do not apply this proposed limit to our analysis.

\section{Constraints from $B$ Decays}

\subsection{Theoretical expressions}

We begin by discussing the constraints which may be obtained from inclusive
semileptonic $B$ decays.  The theoretical treatment of these decays involves a
double expansion in powers of $\alpha_s(m_b)$ and $1/m_b$, employing an
Operator Product Expansion (OPE) and heavy quark symmetry.  From
Ref.~\cite{FLS} we have the expressions for the first two
moments of the hadronic invariant mass spectrum for the process $B\rightarrow
X_c\ell\nu$,
\begin{eqnarray}\label{massivetwo}
   \langle s_H-\overline m_D^2\rangle=&&m_B^2\left[0.051{\alpha_s\over\pi}
   +0.23{\bar\Lambda\over m_B}\left(1+0.43{\alpha_s\over\pi}
   \right)+0.26{\bar\Lambda^2\over m_B^2}
   +1.01{\lambda_1\over m_B^2}-0.31{\lambda_2\over m_B^2}\right],
  \nonumber \\
  \langle (s_H^2-\overline m_D^2)^2\rangle=&&m_B^4\left[0.0053
   {\alpha_s\over\pi}
   +0.067{\bar\Lambda\over m_B}{\alpha_s\over\pi}
   +0.065{\bar\Lambda^2\over m_B^2}-0.14{\lambda_1\over
   m_B^2}\right]\,,
\end{eqnarray}
where $m_\ell=0$, and we have defined the spin-averaged $D$ and $B$ meson
masses,
\begin{eqnarray}\label{massratio}
  \overline m_D\equiv{m_D+3 m_{D^*}\over4}&=&m_c+\bar\Lambda-{\lambda_1\over
  2m_D}+\ldots\simeq 1975\,{\rm MeV}\,,\nonumber\\
  \overline m_B\equiv{m_B+3 m_{B^*}\over4}&=&m_b+\bar\Lambda-{\lambda_1\over
  2m_B}+\ldots\simeq 5313\,{\rm MeV}\,.
\end{eqnarray}
In deriving the expressions (\ref{massivetwo}), we have eliminated
the ratio of pole masses $m_c/m_b$ by instead
writing the heavy quark expansion in terms of $\overline m_D/\overline
m_B$,
\begin{eqnarray}\label{ratiodef}
  {m_c\over m_b}&=&{\overline m_D\over\overline m_B}-{\bar\Lambda\over m_B}
   \left(
  1-{\overline m_D\over \overline m_B}\right)-{\bar\Lambda^2\over m_B^2}\left(
  1-{\overline m_D\over \overline m_B}\right)+{\lambda_1\over 2 m_B m_D}\left(
  1-{\overline m_D^2\over \overline m_B^2}\right) \nonumber\\
  &=&0.372-0.628{\bar\Lambda\over m_B}-0.628{\bar\Lambda^2\over
  m_B^2}+1.16{\lambda_1\over m_B^2}\,.
\end{eqnarray}
Performing a similar substitution in the expression for the semileptonic
decay rate, we
find (for $m_\ell=0$)
\begin{eqnarray}\label{gammaexpr2}
   \Gamma_{\rm s.l.}(B)={G_F^2|V_{cb}|^2m_B^5\over192\pi^3}\,\,0.369
    \bigg[1&&-1.54{\alpha_s\over\pi}-1.65{\bar\Lambda\over m_B}
    \left(1-0.87{\alpha_s\over\pi}\right)\nonumber\\
    &&\mbox{}-0.95{\bar\Lambda^2\over m_B^2}
    -3.18{\lambda_1\over m_B^2}+
    0.02{\lambda_2\over m_B^2}\bigg]\,.
\end{eqnarray}
The advantage of writing $\Gamma_{\rm s.l.}(B)$ and
the moments (\ref{massivetwo}) in this way is that
there is now no hidden dependence on the heavy quark masses;
the coefficients arising at each order in the OPE are
determined by measurable quantities.

The moments of the invariant mass spectrum depend only on
the nonperturbative parameters $\bar\Lambda$, $\lambda_1$
and $\lambda_2$, and on the strong coupling
constant $\alpha_s(m_b)$ at leading order.
Since $\lambda_2(m_b)=0.12\,{\rm GeV}^2$
is known  from the $B$--$B^*$ mass splitting,\footnote{In this paper, we will
neglect the small running
of $\lambda_2(\mu)$ between $m_b$ and $m_c$.} and $\alpha_s(m_b)$
is measured in other  processes, these moments provide direct
information on the unknown hadronic matrix elements $\bar\Lambda$
and $\lambda_1$.
This information may then be inserted into the expression for
$\Gamma_{\rm s.l.}(B)$
to determine $|V_{cb}|$ from the measured decay rate.

\subsection{BLM scale setting for finite $m_c$}\label{blmsec}

We begin by addressing the
question of whether the perturbative corrections to the relation
between the semileptonic decay width and the moments of the hadronic invariant
mass spectrum are in fact well
behaved.  Earlier analyses~\cite{LSW} have indicated that the two-loop
corrections
to $\Gamma_{\rm s.l.}(B)$ are uncomfortably large.  In these analyses, one
computes
that part of the two-loop correction which is proportional
to the first coefficient $\beta_0=11-{2\over3}n_f$ in the QCD beta function,
and from this derives a BLM scale~\cite{BLM} for the process.  One finds the
result
\begin{eqnarray}\label{gammatwoloop}
   \Gamma(B\rightarrow X_c \ell\bar
   \nu)={G_F^2|V_{cb}|^2m_B^5\over192\pi^3}\,\,
   0.369\bigg[1-1.54{\alpha_s(m_b)\over\pi}
   &&-1.43\left({\alpha_s(m_b)\over\pi}\right)^2\beta_0\nonumber\\
   &&-1.65{\bar\Lambda\over m_B}+\dots\bigg]\,,
\end{eqnarray}
which, since $\beta_0\alpha_s(m_b)/\pi\sim 0.6$, leads to a perturbation series
which is quite poorly behaved.  Following the BLM prescription of absorbing
this term into
the ${\cal O} (\alpha_s)$ correction by a change of scale, one finds a low
BLM scale,
$\mu_{\rm BLM}=0.16\,m_b\approx 800\,{\rm MeV}$.\footnote{This scale
arises from taking
$m_c/m_b=0.37$ (see Eq.\ (\ref{ratiodef})),
and differs slightly from the result of Ref.~\cite{LSW}, where
$m_c/m_b$ was taken to be 0.3.}

In Ref.~\cite{FLS} we discussed a similar situation  in the  analysis of the
decay $b\to
u\ell\nu$.   There we considered two perturbation series, neither of which is
particularly
well behaved:
\begin{eqnarray}\label{series}
   \Gamma(B\to X_u\ell\nu)&=&{G_F^2|V_{ub}|^2m_B^5\over192\pi^3}
   \left[1-2.41{\alpha_s(m_b)\over\pi}-2.98\beta_0
   \left({\alpha_s(m_b)\over\pi}
   \right)^2-5{\bar\Lambda\over m_B}+\dots\right]\,,\cr
   \langle s_H\rangle &=& m_B^2
   \left[0.20{\alpha_s(m_b)\over\pi}+0.35\beta_0
   \left({\alpha_s(m_b)\over\pi}
   \right)^2+{7\over10}{\bar\Lambda\over m_B}+\dots\right]\,.
\end{eqnarray}
The BLM scale for $\Gamma(B\to X_u\ell\nu)$ is $\mu_{\rm
BLM}=0.08\,m_b$, while for $\langle s_H\rangle$ it is $\mu_{\rm
BLM}=0.03\,m_b$.
However, both of the expressions (\ref{series}) depend on the nonperturbative
parameter $\bar\Lambda$, which is defined only up to certain arbitrary
conventions\cite{renorms}.  If the poor convergence of the perturbation series
can be absorbed into $\bar\Lambda$, then the large higher-order terms will
be of no consequence, since ultimately $\bar\Lambda$ is eliminated
from relations between physical observables.

In Ref.~\cite{FLS} we investigated whether this might be so by eliminating
$\bar\Lambda$ from the equations (\ref{series}), solving for $\Gamma(B\to
X_u\ell\nu)$ in terms of $\langle s_H\rangle$.  Doing so, we found
\begin{equation}\label{bustrong}
   \Gamma(B\to X_u\ell\nu)={G_F^2|V_{ub}|^2m_B^5\over192\pi^3}
   \left[1-0.98{\alpha_s(m_b)\over\pi}-0.48\beta_0
   \left({\alpha_s(m_b)\over\pi}
   \right)^2-7.14{\langle s_H\rangle\over m_B^2}+\dots\right]\,,
\end{equation}
leading to a much higher BLM scale, $\mu_{\rm BLM}=0.38\,m_b$.  The apparent
convergence of the perturbation series improves considerably under such a
reorganization.

For finite charm quark mass we may perform a similar analysis.  We use
standard techniques~\cite{SmVol} to extract the two-loop term of the form
$\beta_0(\alpha_s/\pi)^2$ which contributes to the first moment $\langle
s_H-\bar
m_D^2\rangle$. The calculation is straightforward but tedious, with the final
integrals performed numerically. We find
\begin{equation}\label{s1twoloop}
   \langle s_H-\overline m_D^2\rangle=m_B^2\left[0.051{\alpha_s(m_b)\over\pi}
   +0.096\left({\alpha_s(m_b)\over\pi}\right)^2\beta_0
   +0.23{\bar\Lambda\over m_B}+\dots\right]\,,
\end{equation}
which again leads to a perturbation series which appears to be badly behaved,
with a very low BLM scale, $\mu_{\rm BLM}=0.02\,m_b$. However, if instead we
use the expression (\ref{s1twoloop}) to eliminate
$\bar\Lambda$ from the semileptonic width (\ref{gammatwoloop}), we obtain
\begin{eqnarray}\label{gammanew}
   \Gamma(B\rightarrow X_c \ell\bar\nu)={G_F^2|V_{cb}|^2m_B^5\over192\pi^3}\,\,
   0.369\bigg[1-1.17{\alpha_s(m_b)\over\pi}
   &&-0.74\left({\alpha_s(m_b)\over\pi}\right)^2\beta_0\nonumber\\
   &&-7.17{\langle s_H-\overline m_D^2\rangle\over m_B^2}+\dots\bigg]\,.
\end{eqnarray}
The two-loop correction in Eq.\ (\ref{gammanew}) has been reduced by a factor
of
almost two compared with that in Eq.\ (\ref{gammatwoloop}).  The one-loop
correction
is reduced as well, so the change in the BLM scale is less dramatic; the new
BLM scale is
$\mu_{BLM}=0.28\,m_b$.
This reorganization of the perturbation series gives us hope that the expansion
$\alpha_s(m_b)/\pi$ is now under control, although, of course, the full
${\cal O}(\alpha_s^2)$ correction remains an important source of uncertainty.

For the second moment of the invariant mass spectrum there is, as in the
massless
case, no such cancelation.  We find, for $\alpha_s(m_b)=0.22$,
\begin{eqnarray}\label{s2twoloop}
   \langle (s_H-\overline m_D^2)^2
   \rangle&=&m_B^4\left[0.0053{\alpha_s(m_b)\over\pi}
   +0.0078\left({\alpha_s(m_b)\over\pi}\right)^2\beta_0
   +0.067{\bar\Lambda\over m_B}{\alpha_s(m_b)\over\pi}+\dots\right]
   \nonumber \\
   &\simeq&m_B^4\left[3.7\times 10^{-4}+3.4\times10^{-4}
   +0.067{\bar\Lambda\over m_B}{\alpha_s(m_b)\over\pi}+\dots\right].
\end{eqnarray}
Since the ${\cal O} (\bar\Lambda)$ term comes with an explicit factor of
$\alpha_s$, substituting
a physical quantity for $\bar\Lambda$ will not
introduce a term
of ${\cal O} (\alpha_s^2\beta_0)$ to cancel the large two-loop correction in
Eq.~(\ref{s2twoloop}).
Therefore, we expect that constraints from the second moment will be more
sensitive to higher
order corrections than those from the first moment and hence less reliable.
Fortunately, the most
useful constraints in the $\bar\Lambda-\lambda_1$ plane will come from the
first moment of
$s_H-\overline m_D^2$.

\subsection{Bounds on $\bar\Lambda$ and $\lambda_1$}

Using the theoretical expressions (\ref{massivetwo}) and experimental
data, we now derive constraints on the nonperturbative parameters $\bar\Lambda$
and
$\lambda_1$.  These quantities dependent on the scheme by which perturbation
theory is defined; the bounds which we will derive are for
$\bar\Lambda$ and $\lambda_1$ at one loop in QCD in the $\overline{\rm MS}$
scheme,
with the renormalization scale $\mu=m_b$.  We make no claim that this is the
``natural'' definition of these quantities, and in any case the scheme
dependence
drops out of
relations between physical observables.  However, although they
are unphysical, it is convenient to retain these
parameters in intermediate stages of calculations, and in order to compare
the values of
$\bar\Lambda$ and $\lambda_1$ obtained from different observables we must
specify some convention for their definition.

In Ref.~\cite{FLS}, we used the known branching ratio of
$B$ mesons to  excited charmed mesons to estimate experimental
lower bounds on $\langle (s_H-\overline m_D^2)^n\rangle$.
This estimate was based on the OPAL  measurement~\cite{OPAL}
of $34\pm7\%$ for the fraction of semileptonic decays to the
states $D_1$ and $D_2^*$.
However, while the sum of the two branching fractions is consistent
with the recent CLEO $90\%$ c.l.~upper limit~\cite{CLEO} of
$30\%$, there appears to be a discrepancy with the
branching fractions to the individual $D_1$ and $D_2^*$ final states.
In Ref.~\cite{FLS} we took the average invariant mass of the
produced $D_1(2420)$ and $D_2^*(2460)$ states to be 2450 GeV.
Here we will assume that only
the lower mass $D_1$ is produced, giving a more conservative
lower limit on $\langle (s_H-\overline m_D^2)^n\rangle$.  We will also take the
$1\sigma$ OPAL lower  limit on the fraction of semileptonic
decays, $27\%$, so as to  be
consistent with the CLEO result.  Doing so, and using the results of
Ref.~\cite{FLS} for the contribution to the moments from the $D$ and $D^*$,
we find the experimental lower limits
\begin{eqnarray}\label{momlimits}
   \langle s_H-\overline m_D^2\rangle_{\rm min} &=& 0.49\,{\rm GeV}^2\,,\cr
   \langle(s_H-\overline m_D^2)^2\rangle_{\rm min} &=& 1.1\,{\rm GeV}^4\,.
\end{eqnarray}
Note that in obtaining these limits we have assumed that no other excited
states
are produced.  It is more realistic to assume that there will also be
production
of the $p$ wave doublet $D_0^*$ and $D_1$, which will raise the average
invariant mass of the final hadronic state.  However, since there is no
experimental information on these states,
we are conservative and do not
include them in our estimates of $\langle (s_H-\overline m_D^2)^n\rangle_{\rm
min}$.

Another observable which depends on $\bar\Lambda$ and $\lambda_1$ is
the ratio of partial widths
\begin{equation}
   R_\tau={\Gamma(B\to X_c\tau\bar\nu)\over\Gamma(B\to X_c e\nu)}\,.
\end{equation}
The theoretical expression for $R_\tau$ also depends on the ratio of masses
$m_\tau/m_b$, both at tree level and in the
nonperturbative~\cite{FLNN,Koyrakh} and
perturbative~\cite{Nir,cjk95} corrections.  As before, $m_\tau/m_b$ may be
re-expanded in terms of the observable $m_\tau/m_B$, yielding the result
\begin{equation}\label{tauratio}
   R_\tau=0.224\left[1+0.24{\alpha_s\over\pi}
   -0.29{\bar\Lambda\over m_B}\left(1-1.33{\alpha_s\over\pi}
   \right)-0.68{\bar\Lambda^2\over m_B^2}
   -3.85{\lambda_1\over m_B^2}-7.54{\lambda_2\over m_B^2}\right]\,.
\end{equation}
At present, there are only data on the average $b$ hadron semitauonic
branching fraction, obtained at LEP, where the identity of
the bottom hadron is not determined.  The experimental result is
\begin{equation}\label{Rexp}
   Br(b\to X_c\tau\bar\nu)=2.75\pm0.48\%\,.
\end{equation}
This differs from $Br(B\rightarrow X_c\tau\bar\nu)$
by contamination from the $B_s$ and $\Lambda_b$.
However, this difference is small
compared with the
experimental uncertainty.  The ratio
of the theoretical expressions for $R_\tau$ in
the $B$ and $\Lambda_b$ sectors is
\begin{equation}
  {R_\tau^B\over R_\tau^{\Lambda_b}}\simeq 1+
  {\cal O}\left({\Lambda_{\rm QCD}^2\over m_B^2}\right)\,.
\end{equation}
Since only about 10\% of $b$ hadrons at LEP are $\Lambda_b$'s, the
effect on $R_\tau$ should be much less than $1\%$.  Hence we use the
measurement
(\ref{Rexp}), along  with ${\rm Br}(B\to X_c e\nu)=10.7 \pm 0.5$
\cite{richbur95a}, to obtain
\begin{equation}
  R_\tau = 0.26\pm0.05\,.
\end{equation}

\begin{figure}
\epsfxsize=13cm
\hfil\epsfbox{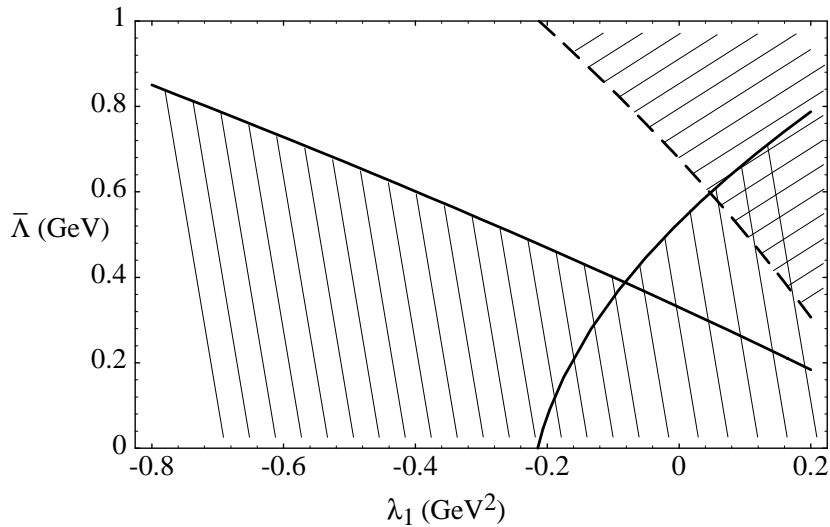}\hfil
\caption{The limits on $\bar\Lambda$ and $\lambda_1$ from data on
semileptonic $B$ decays.
The solid curves are lower limits on $\bar\Lambda$ from the first two moments
of the hadronic invariant mass spectrum, while the dashed curve is a
$1\sigma$ upper limit from the ratio $R_\tau$.  The solid curve on the left
corresponds to the bound from $\langle s_H-\overline m_D^2\rangle$, the one on
the right to the bound from $\langle (s_H-\overline m_D^2)^2\rangle$.}
\label{lambdacurves}
\end{figure}

The comparison of the theoretical predictions (\ref{massivetwo}) and
(\ref{tauratio})
with experiment leads to limits on $\bar\Lambda$ and $\lambda_1$.
The experimental central value for $R_\tau$ yields a curve which is  entirely
inconsistent with the other data, giving a negative value for $\bar\Lambda$.
Therefore in  Fig.~\ref{lambdacurves} we show the curve corresponding to the
$1\sigma$ lower
limit on $R_\tau$, along with the constraints from the moments of the
invariant mass spectrum, where we have taken
$\alpha_s(m_b)=0.22$.
Since the experimental error on  $R_\tau$ is relatively large,
the $2\sigma$ constraint is uninteresting, allowing
all values of $\overline{\Lambda}$ and $\lambda_1$ in
the displayed region of Fig.~\ref{lambdacurves}.  Therefore, at present
we can only conclude that $R_\tau$ favours a negative value of $\lambda_1$.
However, if the experimental uncertainty in $R_\tau$ is reduced in the future,
it may
become an important quantity for constraining $\bar\Lambda$ and
$\lambda_1$.  For now, the most interesting constraints in the
$\bar\Lambda-\lambda_1$
plane come from $\langle s_H-\overline m_D^2\rangle$.

We note that a very similar discussion of the limits on $\bar\Lambda$ and
$\lambda_1$ which may be obtained from $R_\tau$ has been given by
Ligeti and Nir~\cite{LN}.  Our analysis is organized somewhat differently from
theirs in its treatment of experimental masses and errors, leading to results
of a superficially different form, but the physics, and the uncertainties, are
largely the same.

\subsection{Constraints on $|V_{cb}|$}

Leaving aside the weak constraints from $R_\tau$, we now take the information
on
$\bar\Lambda$ and $\lambda_1$ obtained from the analysis of the moments of
$s_H$ and apply it to the extraction of $|V_{cb}|$ from the semileptonic width.
We use the theoretical expression~(\ref{gammaexpr2}), the
experimental semileptonic branching ratio of $10.7\%$
\cite{richbur95a},
the central value $\tau_B=1.60\,$ps for the $B$ lifetime,
and the strong coupling constant $\alpha_s(m_b)=0.22$.
The experimental lower limit on $\langle s_H-\overline m_D^2\rangle$
gives a restriction on the one-loop value of $\bar\Lambda$,
\begin{eqnarray}
   \bar\Lambda&>&\left[0.40-1.15{\alpha_s(m_b)\over\pi}
   -0.07\left({\lambda_1\over0.1\,{\rm GeV}^2}\right)\right]\,{\rm GeV}
   \nonumber \\
   &>&\left[0.32-0.07\left({\lambda_1\over0.1\,{\rm GeV}^2}\right)
   \right]\,{\rm GeV}.
\end{eqnarray}
Including the ${\cal O} (\alpha_s^2\beta_0)$ term and assuming that this
dominates the
two-loop result, we obtain for the two-loop value of $\bar\Lambda$ the
constraint
\begin{eqnarray}
    \bar\Lambda^{(2\,{\rm loop})}&>&\left[0.40-1.15{\alpha_s(m_b)\over\pi}-
    1.88\beta_0\left({\alpha_s(m_b)\over\pi}\right)^2
   -0.07\left({\lambda_1\over0.1\,{\rm GeV}^2}\right)\right]\,{\rm GeV}
   \nonumber \\
    &>&\left[0.25-0.07\left({\lambda_1\over0.1\,{\rm GeV}^2}\right)
   \right]\,{\rm GeV}\,.
\end{eqnarray}
Incorporating the latter bound into our expression for the semileptonic width
and
solving for $|V_{cb}|$, we find
\begin{eqnarray}\label{Vcbresult}
   |V_{cb}| &>& \left[0.038+0.023{\alpha_s(m_b)\over\pi}+
    0.018\beta_0\left({\alpha_s(m_b)\over\pi}\right)^2-2.9\times10^{-4}
   \left({\lambda_1\over0.1\,{\rm GeV}^2}\right)\right]
   \left({\tau_B\over1.60\,{\rm ps}}\right)^{-{1\over2}}\nonumber \\
   &>&\left[0.038+0.0016+0.0006-2.9\times10^{-4}
   \left({\lambda_1\over0.1\,{\rm GeV}^2}\right)\right]
   \left({\tau_B\over1.60\,{\rm ps}}\right)^{-{1\over2}}\nonumber \\
   &>&\left[0.040-2.9\times10^{-4}
   \left({\lambda_1\over0.1\,{\rm GeV}^2}\right)\right]
   \left({\tau_B\over1.60\,{\rm ps}}\right)^{-{1\over2}}.
\end{eqnarray}
In the second line above we display the tree-level, one-loop and (partial)
two-loop contributions to the bound.  As we showed earlier (see
Eq.~(\ref{gammanew})), the perturbation series appears to be well behaved.
We also note that this lower limit on $|V_{cb}|$ is relatively insensitive to
the experimental error on $\langle s_H-\overline m_D^2\rangle$.  If the
semileptonic
production of $D_1$ and $D_2^*$ mesons
is reduced to its $2\sigma$ OPAL lower limit of $20\%$, then the bound on
$\bar\Lambda$ is weakened by $100\,$MeV.
However, we still obtain $|V_{cb}| >
[0.038-2.8\times10^{-4}(\lambda_1/0.1\,{\rm GeV}^2)](\tau_B/1.60\,{\rm
ps})^{-1/2}$.

\subsection{Constraints on $\overline m_b(m_b)$}

Our bounds on $\bar\Lambda$ and $\lambda_1$ may be translated into bounds
on the ${\overline{\rm MS}}$ quark mass $\overline m_b(m_b)$.  However, these
results should be treated with some caution because the
two-loop corrections between
$\overline m_b(m_b)$ and the first two moments of
$s_H-\overline m_D^2$ are quite large, indicating a poorly-behaved perturbation
series.

We define the dimensionless parameter
\begin{equation}
  x_B={m_B-\overline m_b(m_b)\over m_B}={\bar\Lambda\over m_B}+{4\over
  3}{\alpha_s(m_b)
  \over\pi}+1.56\beta_0\left({\alpha_s(m_b)\over\pi}\right)^2+\ldots\,,
\end{equation}
where we keep only the large ${\cal O} \left(\alpha_s^2\beta_0\right)$
contribution to the two-loop term.
We then find
\begin{eqnarray}\label{firstmsbar}
  {\langle s_H-\overline m_D^2\rangle\over
   m_B^2}&=&0.23x_B\left(1-0.52{\alpha_s(m_b)\over\pi}
  +1.13 x_B\right)-0.26{\alpha_s(m_b)\over\pi}
  -0.26\beta_0\left({\alpha_s(m_b)\over\pi}
  \right)^2\nonumber\\ &&\qquad \mbox{}+
  1.13{\lambda_1\over m_B^2}+0.03{\lambda_2\over m_B^2}+\ldots
\end{eqnarray}
for the first moment, and
\begin{eqnarray}\label{secondmsbar}
  {\langle (s_H-\overline m_D)^2\rangle\over
  m_B^4}&=&x_B\left(-0.11{\alpha_s(m_b)\over\pi}
  +0.065 x_B\right)+0.0053 {\alpha_s(m_b)\over\pi}\nonumber\\ &&\qquad\qquad
  \mbox{}+0.0078\beta_0\left({\alpha_s(m_b)\over\pi}
  \right)^2-0.14{\lambda_1\over m_B^2}+\ldots
\end{eqnarray}
for the second.   The ${\cal O}(\alpha_s^2\beta_0)$ corrections are clearly
substantial.
The corresponding BLM scales in Eqs.\ (\ref{firstmsbar}) and
(\ref{secondmsbar})
are $\mu_{\rm BLM}=0.14\,m_b$ and $0.05\,m_b$, respectively, corresponding
to ${\cal O}(\alpha_s^2\beta_0)$ terms which are roughly 60\% and 80\% of the
one-loop term.   Therefore there are likely to be much larger uncalculated
radiative corrections in the relations between the first moment and
$\overline m_b(m_b)$ then between the first moment and the semileptonic $B$
width.

\begin{figure}
\epsfxsize=13cm
\hfil\epsfbox{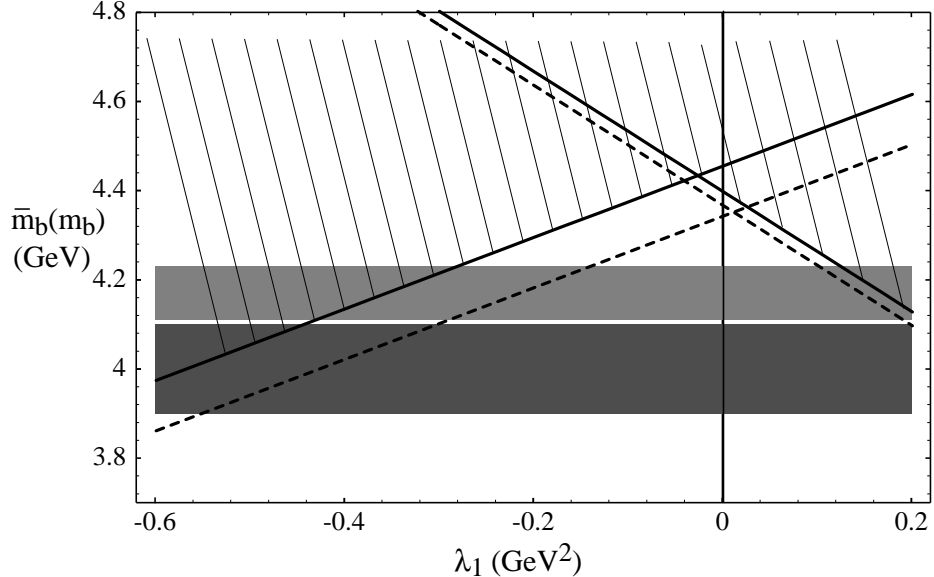}\hfil
\caption{The two-loop  constraints on $\overline
m_b(m_b)$ and
$\lambda_1$ from the first two moments of $s_H-\overline m_D^2$, evaluated at
$\mu=m_b$ (solid lines) and $\mu=m_b/2$ (dashed lines).  The hatched region is
excluded.  Note that the constraints
become more stringent as $\mu$ is lowered.  The two
solid bands are the lattice results from Refs.~\protect{\cite{cgms95,dav95}}
with the quoted uncertainties.
The two-loop calculation includes only terms of order $\alpha_s^2\beta_0$.}
\label{mbconst}
\end{figure}

With these caveats in mind, we combine these results with the experimental
limits~(\ref{momlimits}) to yield the constraints on $\overline m_b(m_b)$ and
$\lambda_1$ shown in
Fig.~\ref{mbconst}.  To illustrate some of the remaining dependence on the
renormalization scale $\mu$, the constraints are plotted for both $\mu=m_b$ and
$\mu=m_b/2$. We also display, with the quoted uncertainties, the results of
two recent lattice extractions of $\overline m_b(m_b)$,
\begin{equation}
  \overline m_b(m_b)=4.17\pm 0.06\,{\rm GeV}~\cite{cgms95}\,,
\end{equation}
and
\begin{equation}
  \overline m_b(m_b)=4.0\pm 0.1\,{\rm GeV}~\cite{dav95}\,.
\end{equation}
We see that our bounds are consistent with these lattice results.
However, it
is much more difficult to compare these limits to the extraction of
the $b$ quark pole mass from
high moments of $\sigma(e^+e^-\to\bar b\, b)$~\cite{vol95},
\begin{equation}\label{vollim}
  m_b=4.827\pm 0.007\,{\rm GeV}\,.
\end{equation}
This is simply because the ambiguity in the renormalization scale introduces a
large uncertainty, of order $\beta_0(\alpha_s/\pi)^2 m_b\sim
200\,$MeV, in the relation of the ``one-loop'' pole mass $m_b^{\rm pole}$ to
$\overline m_b(m_b)$.  For example, if we use $\alpha_s(m_b)$ in the one-loop
relation
\begin{equation}\label{massrel}
  m_b^{\rm pole}\equiv\overline m_b(m_b)\left(1+{4\over 3}{\alpha_s\over\pi}
\right)\,,
\end{equation}
then Eq.~(\ref{vollim}) yields the value $\overline m_b(m_b)=4.42$ GeV,
somewhat higher than the lattice results and barely consistent with our
analysis of the moments. Furthermore, such a value certainly would be
inconsistent with the combination of the moments analysis and the QCD sum rules
extraction
of $\lambda_1=-0.6\,{\rm GeV}^2$.  On the other hand, in Ref.~\cite{vol95} it
was
argued that the
natural scale for matching onto the non-relativistic theory is $\mu\sim
0.63\,m_b$.  Taking this lower scale, we find $\overline
m_b(m_b)=4.34\,$ GeV, still
somewhat higher
than the lattice calculations and still inconsistent with a large negative
$\lambda_1$.
Of course, using an even lower scale in the one loop
relation between $m_b^{\rm pole}$ and $\overline m_b(m_b)$ would
lower the extracted value of $\overline m_b(m_b)$ even further.
While this scale ambiguity is formally of higher order in
$\alpha_s$ than our calculation, we see that numerically it is quite
significant.  Without a higher loop calculation of $m_b^{\rm pole}$ from QCD
sum rules, it is
impossible to determine whether or not this extraction of
$m_b^{\rm pole}$ is consistent with the other constraints.

\section{D Meson Decays}

In Ref.~\cite{bds94} it was argued that the semileptonic decays of
charmed mesons are not well described in the heavy quark expansion,
since the value of $m_c$ which is required to fit the observed
semileptonic $D$ decay rate lies significantly above 1.4 GeV.  This is
the upper limit suggested by combining the value of of $m_b$ extracted
from the $\Upsilon$ spectrum~\cite{vol95} and the large negative value
of  $\lambda_1$ found from certain QCD sum rules~\cite{BB93a}.
However, we believe that this argument should be reconsidered in light
of the uncertainty inherent in relating the pole mass derived in
Ref.~\cite{vol95} with physical quantities. Indeed, we will find that
the values of $\lambda_1$ and $\bar\Lambda$ implied by the semileptonic
decay rate and the first moments of the $D$ and $D_s$ invariant mass
spectra are in reasonable agreement with the limits from the
corresponding observables in the bottom sector.

The theoretical analysis in the $D$ sector proceeds as before. Since
$m_s$ is of order $\Lambda_{\rm QCD}$, for consistency we keep only terms
of order $m_s^2/m_c^2$ in the theoretical expressions for $\langle
s_H^n\rangle$ and $\Gamma(D\to X_s e\bar\nu)$.  As we are neglecting
terms of order $\alpha_s\lambda_1$ and $\alpha_s\lambda_2$, we also omit
terms of order $\alpha_s m_s^2$ but keep logarithmically enhanced terms
of order $\alpha_s m_s^2\ln (m_s^2/m_c^2)$. Thus we find for the Cabibbo
allowed semileptonic width,
\begin{eqnarray}\label{Drate}
   \Gamma(D\rightarrow X_s e\bar\nu)&
   =&{G_F^2 m_c^5|V_{cs}|^2\over 192\pi^3}
   \left[1+\left({25\over 6}-{2\over 3}\pi^2\right){\alpha_s\over\pi}
   -8 {m_s^2\over m_D^2}\left(1-2{\alpha_s\over\pi}
   \ln{m_s^2\over m_c^2}\right)
   +{\lambda_1-9\lambda_2\over 2 m_D^2}\right] \nonumber
   \\ &=&{G_F^2 m_c^5|V_{cs}|^2\over 192\pi^3}
   \left[1+\left({25\over 6}-{2\over 3}\pi^2\right){\alpha_s\over\pi}
   -8  {\overline m_s^2(m_c)\over m_D^2}+{\lambda_1-9\lambda_2\over
   2 m_D^2}\right],
\end{eqnarray}
and for the first moment
\begin{eqnarray}\label{Dmom}
   \langle s_H\rangle_D&=&
   m_s^2\left(1-2{\alpha_s\over\pi}\ln{m_s^2\over m_c^2}\right)+
   m_D^2\left[{91\over 450}{\alpha_s\over\pi}+{7\bar\Lambda\over 10 m_D}
   \left(1-{227\over 630}{\alpha_s\over\pi}\right)+{3\over 10 m_D^2}
   \left(\bar\Lambda^2+\lambda_1-\lambda_2\right)\right]\nonumber\\
   &=&\overline m_s^2(m_c)+
   m_D^2\left[{91\over 450}{\alpha_s\over\pi}+{7\bar\Lambda\over 10 m_D}
   \left(1-{227\over 630}{\alpha_s\over\pi}\right)+{3\over 10 m_D^2}
   \left(\bar\Lambda^2+\lambda_1-\lambda_2\right)\right].
\end{eqnarray}
Note that the large infrared logarithms of the pole mass $m_s$ may be
absorbed naturally into the $\overline{\rm MS}$ mass renormalized at
$m_c$, $\overline m_s(m_c)$.  As one would expect, the individual terms in
the perturbative expansion, which arises from an operator product
expansion performed at the scale $\mu=m_c$, remain insensitive to physics
below $m_c$.

In addition to the usual nonperturbative parameters, the theoretical
expressions for the decay rate and moments also depend on the strange
quark mass $\overline m_s(m_c)$. We use the range
\begin{equation}\label{smass}
  100\,{\rm MeV}<\overline m_s(1\,{\rm GeV})<300\,{\rm MeV}\,,
\end{equation}
given by the Particle Data Group \cite{PDG}
(which changes only slightly when evolved from $\mu=1\,{\rm GeV}$ to
$\mu=m_c$).
Finally, note that
we have not re-expanded the $m_c^5$ term appearing in the semileptonic width
(\ref{Drate})
in terms of the meson
mass $\overline m_D$.  This is because here we have no analogue of the
expansion
(\ref{ratiodef}), since the strange quark is not heavy.  As a result, the
parameter $\bar\Lambda$ does not appear explicitly in Eq.~(\ref{Drate}).
Instead, we will solve Eq.~(\ref{Drate}) directly for $m_c$, linearize in
$\lambda_1$, and then use the
heavy quark expansion (\ref{massratio}) to relate $m_c$ to $\bar\Lambda$ and
$\lambda_1$.
\begin{figure}
\epsfxsize=13cm
\hfil\epsfbox{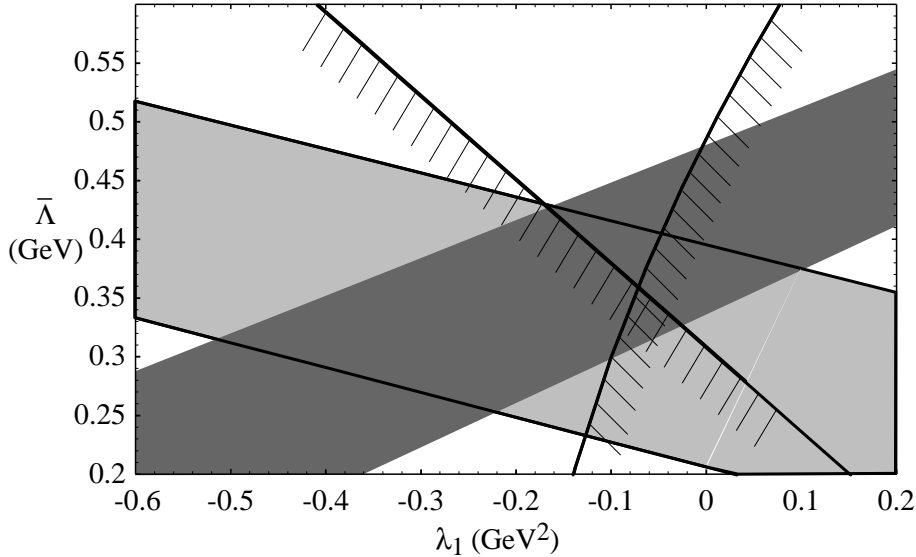}\hfil
\caption{The restrictions on $\bar\Lambda$ and
$\lambda_1$ from the semileptonic $D$ decay rate (darker shaded region)
and from $\langle s_H\rangle$ in $D$ decay
(lighter shaded region).  We also show the lower bounds on
$\bar\Lambda$ from the moments of the $B$ decay spectrum
(cross-hatched black lines).}
\label{charml1lbar}
\end{figure}

The inclusive semielectronic $D$ branching fraction recently has been
measured to be~\cite{cleoD0}
\begin{equation}
  Br(D^0\to Xe^+\nu)=[6.64\pm0.18(stat.)\pm0.29(syst.)]\%\,.
\end{equation}
The same CLEO analysis indicates that the Cabibbo-allowed inclusive
rate is saturated, within errors, by the exclusive modes
$D^0\to(K^-,K^{*-})\,e^+\nu$, with stringent upper limits reported on the
channels $D^0\to(K_1^-(1270),K^{*-}(1430))\,e^+\nu$.  The data on
these decays yield for the first moment of the $D\to X_s\ell^+\nu$
invariant mass spectrum the value
\begin{equation}
   \langle s_H\rangle_D = (0.49\pm0.03)\,{\rm GeV}^2\,,
\end{equation}
where we have added errors in quadrature and neglected the widths
of the excited $K$ mesons.

The result of this analysis is the set of constraints displayed in
Fig.~\ref{charml1lbar}, in which we also show the limits obtained
earlier from $B$ decays. Both theoretical and experimental
uncertainties are included in the displayed bands.  In fact, the
dominant uncertainties are theoretical, and have two distinct sources.
First, there is the uncertainty in the strange quark mass
(\ref{smass}).  Second, there is the effect of uncomputed terms in the
mass expansion of order $1/m_c^3$, which will be more substantial than
in bottom decays.  The theoretical analysis at order $1/m_c^3$ is
quite complex and involves a number of new nonperturbative parameters,
so we do not attempt to include these terms
systematically.\footnote{In Ref. \cite{bds94} the $1/m_c^3$
corrections to the semileptonic decay rate were estimated using the
factorization hypothesis and other arguments, and found to be small.}
Instead, we obtain a minimal estimate of the size of the uncertainty
arising from these effects by extracting the bounds from
$\Gamma(D\rightarrow X_s\ell^+\nu)$ in two ways: on the one hand, by
solving directly for the width in terms of the charm quark pole mass,
and on the other, by proceeding through the intermediate step of
calculating the ``decay mass'' $m_c^\Gamma$~\cite{FLS}. These two
procedures, which are formally the same only up to order $1/m_c^2$,
yield bounds on $\bar\Lambda$ which differ by approximately $70\,{\rm
MeV}$.  It would be hard to argue convincingly that $1/m_c^3$ effects
were intrinsically smaller than this.  For the analysis of $\langle
s_H\rangle_D$, we employ the simpler procedure of including a term
$n(0.5\,{\rm GeV}/m_D)^3$ in the theoretical expression
(\ref{Dmom}) and varying $n$ between $-1$ and 1.

Note that the constraints from the
charm sector are compatible with those
derived from bottom decays, although they do not appear to agree with
the lattice extractions of$m_b$ shown in Fig.~\ref{mbconst}.  However,
Fig.~\ref{charml1lbar} must be interpreted with caution, because of
the possible presence of large higher loop effects.  As discussed in Section
IIC, each curve in Fig.~\ref{charml1lbar} may be interpreted as giving a
constraint on $\bar\Lambda$ derived from a particular process.  A reliable
comparison between the constraints derived from two different physical
quantities requires that when one is expressed in terms of the other, the
perturbative expansion in $\alpha_s$ be well behaved.  For example, we
demonstrated earlier that
although there are large two-loop corrections to the theoretical
expressions for $\Gamma_{\rm s.l.}$ and $\langle s_H\rangle$ for $B$ or $D$
decays alone, these corrections partially cancel out when $\bar\Lambda$ is
eliminated and one of these physical quantities is expressed in terms of the
other.  Unfortunately, this requirement may no longer be satisfied when
expressions from the bottom and charm sectors are compared with each other.
For example, writing $\langle s_H-\overline m_D^2\rangle$ for $B$ decays in
terms of $\langle s_H\rangle$ for $D$ decays, we find
\begin{equation}
   \langle s_H-\overline m_D^2\rangle = m_B^2\left[0.028{\alpha_s(m_c)\over\pi}
   +0.029\beta_0\left({\alpha_s(m_c)\over\pi}\right)^2
   +0.12{\langle s_H\rangle-\overline m_s^2(m_c)\over m_D^2}\right]\,,
\end{equation}
resulting in a pertubative expansion which does not appear to converge well.
Thus while the rough consistency of the
constraints from $B$ and $D$ decays is encouraging, it may not be
particularly significant.  Still, it is amusing to note that if we were to take
the combined constraints as legitimate, then we would conclude that
$\lambda_1$ is small and negative, of order $-0.1$ GeV$^2$. Hence it would
make a negligible contribution to most observables, including $|V_{cb}|$.

We do not compare the results from the charm sector with
$\overline m_c(m_c)$, which may be extracted from the lattice measurement
of $\overline m_b(m_b)$ using the heavy quark mass relations.  This is
because the radiative corrections between these quantities are so large
that perturbation theory appears uncontrolled, making it difficult to
conclude whether or not the regions in the $\bar\Lambda-\lambda_1$ plane
indicated by the different observables are consistent with each other.

A better test of duality in charm decays comes from comparing $D$
and $D_s$ decays.  There is only an upper bound on the semileptonic branching
ratio of the $D_s$~\cite{PDG},
\begin{equation}
  Br(D_s\rightarrow Xe^+\nu)<20\%\,,
\end{equation}
which is not strong enough to provide an interesting constraint from the
semileptonic decay rate.
The observed exclusive semileptonic modes are $D_s\to
(\eta,\eta',\phi)\,\ell^+\nu$, for which CLEO has recently reported
relative branching ratios~\cite{cleoDs}. Adding the reported systematic and
statistical errors in quadrature, we estimate a
value for the first moment of the invariant mass spectrum,
\begin{equation}
   \langle s_H\rangle_{D_s} = (0.68\pm0.03)\,{\rm GeV}^2\,.
\end{equation}
We have assumed that decays to $\eta$, $\eta'$ and $\phi$ saturate the
inclusive semileptonic rate, an approximation for which we do not assign an
error.  We merely note that
since semileptonic $D^0$ decays are known to be saturated by
relatively few exclusive modes, the error due to this
approximation for $D_s$ might well be small.  In light of this
uncertainty, the upper limit on $\langle s_H\rangle_{D_s}$ is perhaps
somewhat less firm than the lower limit.

The theoretical prediction for $\langle s_H\rangle_{D_s}$ is given by
the expression (\ref{Dmom}) for $\langle s_H\rangle_D$, with the
replacements of $m_D$ by $m_{D_s}$ and $\bar\Lambda$ and $\lambda_i$
by their strange counterparts $\bar\Lambda_s$ and $\lambda_{is}$.
They are related to $\bar\Lambda$ and $\lambda_i$ by
\begin{eqnarray}\label{lamdiffs}
   \bar\Lambda_s-\bar\Lambda &=& {\overline m_B
   (\overline m_{B_s}-\overline m_B) -
   \overline m_D(\overline m_{D_s}-\overline m_D)\over \overline m_B-
   \overline m_D}\approx95\,{\rm
   MeV}\,,
   \nonumber\\
   \lambda_{1s}-\lambda_1 &=& {2\overline m_B\overline m_D\over
   \overline m_B-
   \overline m_D}\left[
   (\overline m_{B_s}-\overline m_B) - (\overline m_{D_s}-
   \overline m_D)\right]
   \approx-0.02\,{\rm GeV}^2\,,\\
   \lambda_{2s}-\lambda_2 &=&
   {1\over4}\left[\left(m_{D_s^*}^2-m_{D_s}^2\right)-
   \left(m_{D^*}^2-m_{D}^2\right)\right]\approx-0.01\,{\rm GeV}^2
   \nonumber\,,
\end{eqnarray}
up to corrections to the mass expansions of relative order $1/m_{b,c}^3$.

The theoretical expression for $ \Delta\langle s_H\rangle=\langle
s_H\rangle_{D_s}-\langle s_H\rangle_D$ is particularly well behaved,
since it vanishes up to
$SU(3)$ violating effects.  Expanding in powers of $1/m_D$, we find
\begin{eqnarray}\label{s1diff}
   \Delta\langle s_H\rangle=
   m_D(\bar\Lambda_s-\bar\Lambda)&&\left[{7\over10}+{137\over900}
   {\alpha_s(m_c)\over\pi}\right]
   +{13\over10}\bar\Lambda(\bar\Lambda_s-\bar\Lambda)\nonumber\\
   &&\mbox{}+(\bar\Lambda_s-\bar\Lambda)^2
   +{3\over10}(\lambda_{1s}-\lambda_1)
   -{3\over10}(\lambda_{2s}-\lambda_2)\,.
\end{eqnarray}
The large radiative corrections to $\langle s_H\rangle$ almost precisely
cancel in the difference.

The leading corrections to this result likely come
from the $SU(3)$ violating Cabibbo-allowed annihilation channel in $D_s$
decays.
Although this channel typically has small $s_H$, it contributes to $\langle
s_H\rangle_{D_s}$ primarily through its effect on the total semileptonic rate.
For small
lepton mass, helicity conservation suppresses the purely leptonic decay
$D_s\to\ell\nu$ in favor of processes in which at least one gluon is emitted.
Using vacuum saturation to guess the relevant hadronic matrix elements, one
obtains a na\"\i ve estimate of $4\pi\alpha_s(m_c) f_D^2/m_D^2$ for the
fractional correction to the $D_s$ semileptonic decay width\footnote{
The authors of Ref.~\cite{bds94} have calculated a related quantity of the same
order, namely  the contribution of ``penguin-type'' annihilation processes to
the semileptonic decay of the $D$.  They obtain the na\"\i ve estimate above,
multipled by a coefficient $\case89(\log(m_c/\mu)+1/3)\approx1.3$ for
$\mu\approx500\,$MeV. }.  With
$f_D\approx200\,$MeV and $\alpha_s(m_c)\approx 0.32$, this corresponds to
an increase of order 5\% in the semileptonic width of the $D_s$, or a decrease
of
$\langle s_H\rangle$ by roughly the same percentage.
Since the error in this estimate due to the use of vacuum saturation
is unknown, we will take the effects of weak annihilation into account by
decreasing the predicting for $\langle s_H\rangle_{D_s}$ by 5\%, and assigning
an
additional error of $\pm 5\%$, or 100\% of the na\i\" ve estimate of the
$1/m_D^3$
correction, to the prediction of the differences in the moments.


With the known values (\ref{lamdiffs}) of
$\bar\Lambda_s-\bar\Lambda$ and $\lambda_{is}-\lambda_i$, the expression
(\ref{s1diff}) reduces to $\Delta\langle
s_H\rangle=[0.136+0.121(\bar\Lambda/1\,{\rm
GeV})]\,{\rm GeV}^2$.  For an analysis entirely within the charm system,
it is appropriate to estimate $\bar\Lambda$ from the overlap
of the shaded bands in Fig.~\ref{charml1lbar}, from which we obtain
$0.25\,{\rm GeV}<\bar\Lambda<0.45\,{\rm GeV}$.  Including the estimated
contribution and uncertainty from
the annihilation channel in $D_s$ decay, this range leads to
$\Delta\langle s_H\rangle=(0.14\pm0.04)\,{\rm GeV}^2$, which we may
combine with the measured value of $\langle s_H\rangle_D$ to obtain
the theoretical prediction
\begin{equation}
  \langle s_H\rangle_{D_s} = (0.64\pm0.04)\,{\rm GeV}^2\,.
\end{equation}
This value agrees quite well with the experimental result
$(0.68\pm0.03)\,{\rm GeV}^2$.

Indeed, we view the successful prediction of $\langle
s_H\rangle_{D_s}$ as evidence for the applicability of parton-hadron
duality to inclusive semileptonic charm decays.  The prediction of this
moment is on a firmer theoretical footing than other quantities in the
charm system, since the large radiative and
power corrections up to ${\cal O}(1/m_c^2)$ cancel out of the difference of the
moments. This feature distinguishes this test of duality from those in which
a comparison is made to the bottom system, while its relative
insensitivity to the value of $\bar\Lambda$ makes it
more stringent than the comparison of the first moment to the total $D$
decay width. At the same time, duality is satisfied in a nontrivial way,
as the exclusive final states in $D$ and $D_s$ decays are entirely
different.  We are encouraged by this success of the theoretical analysis.

\section{Summary}

In this paper we have explored the constraints on the nonperturbative
parameters $\bar\Lambda$ and $\lambda_1$ which are obtained from
semileptonic $B$ and $D$ decays.  We have found that independent analyses
of the bottom and charm systems yield limits which are consistent with
one another.  Taken together, they imply values of the order
$\bar\Lambda\sim450\,$MeV (at one loop) and $\lambda_1\sim-0.1\,{\rm
GeV}^2$. Whether or not one chooses to trust the numerical results of the
charm analysis, we see no evidence that parton-hadron duality fails in
these decays.  On the contrary, our discussion of the difference of the
first moments in $D$ and $D_s$ decays leads us to quite the opposite
conclusion:  in at least one nontrivial case, duality works well for
charm.

A primary motivation for investigating inclusive decays is to extract the CKM
matrix element $|V_{cb}|$ with high precision.  Our analysis
yields the lower limit $|V_{cb}| >
[0.040-2.9\times10^{-4}(\lambda_1/0.1\,{\rm GeV}^2)](\tau_B/1.60\,{\rm
ps})^{-1/2}$ using the current measurement of the branching fraction to excited
$D$ meson states in $B$ decays.  This is consistent
with  the value of $|V_{cb}|$ obtained from exclusive $B$
decays~\cite{Nconf}.
We have bolstered our theoretical analysis with a partial treatment of two-loop
corrections to this bound, performing a BLM scale setting analysis which
indicates that the relevant
perturbation series is reasonably well behaved.

\acknowledgements

It is a pleasure to thank Aida El-Khadra, Zoltan Ligeti and Mark Wise for
helpful conversations, and Arne Freyberger for bringing
Refs.~\cite{cleoD0} and \cite{cleoDs} to our attention.  This work was
supported by the United States Department of Energy under Grant
No.~DE-FG02-91ER40682, by the United  States National Science Foundation
under Grant No.~PHY-9404057, and by the Natural Sciences and Engineering
Research Council of Canada.  A.F.~acknowledges additional support from
the United States National Science Foundation for National Young
Investigator Award No.~PHY-9457916, the United States Department of
Energy  for Outstanding Junior Investigator Award No.~DE-FG02-94ER40869
and the Alfred P.~Sloan Foundation.  M.J.S. acknowledges additional
support from the United States Department of Energy for Outstanding
Junior Investigator Award No.~DE-FG02-91ER40682.

\end{document}